# A Vision of Interdisciplinary Education: Students' Reasoning about "High-Energy Bonds" and ATP

Benjamin W. Dreyfus, Vashti Sawtelle, Chandra Turpen, Julia Gouvea, and Edward F. Redish

*Department of Physics, University of Maryland, College Park, MD 20742*

***Abstract.*** As interdisciplinary courses are developed, instructors and researchers have to grapple with questions of how students should make connections across disciplines. We explore the issue of interdisciplinary reconciliation (IDR): how students reconcile seemingly contradictory ideas from different disciplines. While IDR has elements in common with other frameworks for the reconciliation of ideas across contexts, it differs in that each disciplinary idea is considered canonically correct within its own discipline. The setting for the research is an introductory physics course for biology majors that seeks to build greater interdisciplinary coherence and therefore includes biologically relevant topics such as ATP and chemical bond energy. In our case-study data, students grapple with the apparent contradiction between the energy released when the phosphate bond in ATP is broken and the idea that an energy input is required to break a bond. We see students justifying context-dependent modeling choices, showing nuance in articulating how system choices may be related to disciplinary problems of interest. This represents a desired endpoint of IDR, in which students can build coherent connections between concepts from different disciplines while understanding each concept in its own disciplinary context. Our case study also illustrates elements of the instructional environment that play roles in the process of IDR.

## I. INTRODUCTION

It is well-established that physics students can compartmentalize their understanding of the physical world.[1,2] The ideas about physical phenomena and about the nature of knowledge that students bring to bear in physics class settings are often different from the ideas that the same students bring to bear in "everyday" settings. Previous work has focused on developing opportunities for students to reconcile canonical physics concepts with their everyday experience.[3–6] This does not mean learning to discount everyday intuitions, but rather, learning to build coherent connections between the "physics" domain and the "everyday" domain.

In this paper, we turn our attention to a related yet distinct type of compartmentalization, into compartments that have been referred to as "disciplinary silos." Students take physics, biology, and chemistry courses, but rarely have opportunities to bring the ideas of each discipline into direct contact, and disciplinary experts often have limited contact with the other science disciplines. In an age of increased emphasis on interdisciplinary connections among the sciences, we seek to understand the reconciliation of ideas from different science disciplines, with an eye toward clarifying the goals of interdisciplinary science education. While related, interdisciplinary reconciliation is qualitatively different from the reconciliation between "physics" and "everyday" ideas, because it involves reconciling multiple sets of "expert" scientific ideas.

Our investigation has a theoretical goal and draws heavily on case-study data of an introductory physics course for undergraduate life science students, in the context of reasoning about energy and ATP (adenosine triphosphate). We pursue one central research question: How can we characterize interdisciplinary reconciliation in the context of existing frameworks for reconciliation of ideas? In this paper, we explore two specific subquestions in the context of learning about chemical energy as a way to address the more general question: 1) What does successful interdisciplinary reconciliation look like in the context of energy? 2) When biology students encounter ATP in a physics course, how do they negotiate disciplinary differences between biology and physics in this instructional context?

## II. THEORETICAL FRAMEWORK

The guiding theoretical perspective for our analysis is the resources framework, which we now briefly overview. The consensus view in physics education research is constructivism: the idea that learners are not blank slates, but all new knowledge is built on existing knowledge.[7] In order to understand how students learn, PER across the board strongly focuses on understanding the ideas they enter with. However, there are two major ways of characterizing these ideas. Under the misconceptions model [8–10], students possess strongly held, stable, and unitary beliefs, which differ from expert conceptions. That is, if a student holds a misconception, we would expect that student to exhibit that misconception consistently across multiple contexts. In this model, the goal of instruction is to confront and replace misconceptions.

In contrast, the resources framework [11–14] sees students' knowledge as more dynamic, with the possibility of being fragmented. Rather than a single coherent theory that differs from expert understanding, students possess a variety of resources that can be activated differentially in different contexts. For example, a student might apply the concept "motion is caused by a force" in some circumstances but not others.[12] Students might access particular resources in ways that lead them to incorrect conclusions. The goal of instruction is not to eliminate these resources, but to help students use their resources productively and refine their sense of when those resources are most useful. Resources can be both conceptual (resources for understanding physical phenomena) and epistemological [13,15,16] (beliefs about the nature of knowledge and learning). The resources framework also allows for the possibility of stable patterns of resource activation. However, in this framework, stability is one possible description of a set of resources rather than the default assumption, and multiple stabilities can coexist.[17]

The resources framework is based in physics education research but is spreading into other science disciplines,[18,19] and we extend it here to interdisciplinary reasoning. Our analysis involves concepts (in this case, chemical bond energy) that students encounter in multiple contexts associated with multiple disciplines, and so we draw on research on the context dependence of student reasoning. Our objective is to understand how ideas from different disciplines are coordinated in a new context, a phenomenon that falls under the broad class of phenomena often described as "transfer." Hammer et al. [20] argue that transfer phenomena can be understood as the context-dependent activation of cognitive resources. What looks like transferring ideas from one context to another is the activation of similar sets of resources through the generation of similar framings, across different contexts.

Framing is a concept from sociolinguistics [21–23] that describes an individual's expectations or interpretation of "What kinds of knowledge or approaches are appropriate here?"[24] Hammer et al. use the concept of framing to refer to the activation of locally coherent sets of resources. Student reasoning influences and is influenced by the context [25]; this leads to an understanding of framing as emergent from the interaction between the student and the context. Along these lines, Engle et al. define framing as "the metacommunicative act of characterizing what is happening in a given context and how different people are participating in it."[26] This definition of framing gives emphasis to both the physical setting and the social interactions that build up reality in a moment.

As we discuss in greater depth in section VI, the framework of context-dependent activation of resources is relevant to understanding reasoning across disciplines because disciplinary contexts influence (and are influenced by) the conceptual and epistemological resources that students draw on. Furthermore, in addition to the disciplinary context, there are aspects of the instructional context (messages from the instructor that suggest how following messages should be framed, and other elements of the "hidden curriculum") that may contribute to students' framing. Therefore, we also highlight those aspects in our data in order to present a more complete picture of the context in which the reasoning takes place.

# III. SETTING FOR THE CASE STUDY

## A. Energy is an ideal context for studying interdisciplinary reconciliation

There have been many calls for interdisciplinary science education in recent history [27–29] and attempts at integrating the disciplines in the last several decades [30]. However, bringing the disciplines together in a meaningful way is not a trivial process.[31,32] At the University of Maryland, we have been involved in creating an Introductory Physics for the Life Sciences course [33], which at its core attempts to aid students in building connections across the disciplines of physics and biology. One area of focus for creating these connections resides in topical areas that span the disciplines such as energy, thermodynamics, and light (i.e., constructs that are central to each discipline independently). However, we start from the perspective that overlapping content topics alone are insufficient for making meaningful interdisciplinary connections. It is also necessary to attend to how knowledge is structured in and among the disciplines by instructors and by students. A number of other physics courses [34–37] and curricula [38,39] have incorporated strong connections to biology content, and research on some of these courses has evaluated students' conceptual understanding and attitudes (often through assessments developed for conventional physics courses). Still, research that explicitly addresses how students connect ideas from multiple disciplines in those courses is limited. This paper is situated in the context of energy, one of these cross-disciplinary topics, and uses this context for a broader examination of interdisciplinary science education and the reconciliation of concepts between physics and biology.

Understanding the role of energy in biological processes requires understanding ATP (adenosine triphosphate), a molecule that biology students know as "the energy currency of the cell." However, the treatment of energy in the traditional introductory physics curriculum (including the courses taken by most biology students) focuses on mechanical energy, and does not make a clear connection to the energy transformations most relevant to biological systems at the cellular and molecular levels.[40] Developing a physics curriculum for life sciences students that is intended to build cross-disciplinary coherence requires engaging with energy concepts as they are understood and leveraged in biology and chemistry. A major component of supporting this cross-disciplinary coherence requires attending to the energy associated with chemical bonds, and especially ATP.

ATP is produced during cellular respiration and photosynthesis. In the ATP hydrolysis reaction, which takes place in aqueous solution within the cell, a bond is broken to remove the terminal phosphate group from the ATP molecule, leaving ADP (adenosine diphosphate). Breaking this bond (like any bond) requires an input of energy. Both products (ADP and inorganic phosphate) form other bonds as a result of their interaction with water. These new stronger bonds are associated with a greater total bond energy (equivalently, they are represented by a deeper potential well), resulting in a net release of energy. [1] This energy is used to power various cellular processes, by coupling ATP hydrolysis to other reactions. As a shorthand, many biology texts and instructors refer to the phosphate bond in ATP as a "high-energy bond."[41] This terminology may be understood to imply that there is energy "in" this bond that is released when the bond is broken, even though the breaking of this bond itself is not what releases the energy.

Students' conceptual difficulties with ATP and bond energy are well-documented in the biology and chemistry education literatures. In biology, Novick and Gayford both write about student confusion about "energy stored in bonds" and the misleading terminology of "high-energy bonds," particularly in regard to ATP.[42,43] Storey identifies biology textbooks as perpetuating this confusion.[44] In chemistry, Boo documents students' "alternative conception" that bond making requires energy input (even in non-biological reactions), as an alternative to the idea that bond making releases energy.[45] Galley also documents "exothermic bond breaking"[46] in student reasoning, as we will discuss at greater length in sections IV and V. Teichert and Stacy show that students (when discussing ATP) can simultaneously express the idea that energy is released when a bond is formed, and that energy is released when a bond is broken.[47]

This literature does not clearly establish what good interdisciplinary reasoning should look like in the context of ATP. In our approach, our understanding of "good" reasoning emerged from a careful examination and articulation of exemplary student reasoning, rather than having experts set *a priori* metrics of success. This paper further explores conceptions about ATP, presents evidence of students' reconciliation of these ideas in a manner that may be unique to interdisciplinary concepts, and explores how an interdisciplinary instructional context supported this reconciliation.

## B. Course setting

The context of this study was the pilot year of a new introductory physics course[2] for undergraduate biology students.[33] The course is part of the National Experiment in Undergraduate Science Education (NEXUS), a project that is producing competency-based curricula for life science students.[48] It represents the results of an interdisciplinary collaboration [32] bringing together perspectives from physics, biology, biophysics, chemistry, and education research. This course is unusual in that biology and chemistry are required as prerequisites, and students are therefore expected and encouraged to draw on their knowledge from these other disciplines. The course spent substantially more time on energy and thermodynamics than the typical introductory physics course, because these topics are also central to chemistry and biology, and they provide opportunities to build coherence across the disciplines.

Structurally, the course ran as a typical introductory physics course at the university level with three 50-minute lectures per week, accompanied by one 2-hour lab section and one 50-minute discussion section. In contrast to a typical introductory physics course, the class meetings and discussion sections involved extensive group problem-solving tasks that were designed to build connections between chemistry, biology, and physics. In this first pilot year, approximately 20 students were enrolled in the course each semester. One of the authors (Redish) served as the instructor in the course.

Our previous research [49] shows that some students perceive a disconnect between energy in physics and energy in biology, even to the point of thinking about energy in the two disciplines as two separate entities (related only by analogy). One student we interviewed saw this distinction as corresponding to spatial scale, with physics primarily concerned with mechanical (kinetic and potential) energy at the macroscopic scale, and biology concerned with chemical energy at the cellular and molecular scales. Other recent work [50,51] has contrasted the curricular treatment of energy in physics and biology using data from curricula and faculty.

[1] In this case, the qualitative description at this level of detail would be identical if we were discussing Gibbs free energy (which is often the more relevant quantity in many biochemical contexts) rather than energy. For the remainder of the paper, we talk only about energy and not free energy, because the distinction is immaterial here (though we are aware that this distinction is significant in other situations).

[2] See http://nexusphysics.umd.edu

To bridge these various uses of energy, our course included an extensive thread on chemical bond energy [52], emphasizing that the energy associated with chemical bonds is potential and kinetic energy and is included in the overall conservation of energy. The course readings developed the Lennard-Jones potential (mostly qualitatively) as a way to describe the chemical bond in terms of electric potential energy and other constructs "native" to physics courses. Students were given a series of tasks in which they were to model chemical bonds with potential energy graphs (displaying potential energy as a function of position), and to use reasoning similar to conventional conservation-of-mechanical-energy problems. (One homework problem paired a question about interacting atoms with a question about a skateboarder skating down a hill.) Students also used computer simulations from the CLUE curriculum [53] that illustrated the formation of bonds using graphical representations of potential energy. This model of chemical bonds was intended to provide a stronger conceptual foundation for the principle that breaking a bond requires an energy input (and conversely, that forming a bond releases energy), by recognizing that climbing out of a potential well requires an input of energy and represents the breaking of a bond.

# IV. CASE STUDY METHODOLOGY

## A. Data collection

This paper explores a case of interdisciplinary reconciliation in the context of ATP and bond energy through four complementary data sources: 1) quantitative student response data from a multiple-choice quiz question to obtain a baseline for the class as a whole, 2) qualitative data from interviews to examine individual students' thinking in greater detail, 3) in-class video data from the day the quiz was handed back to illustrate how students and the instructor framed the task in that moment, and 4) a capstone essay exam question to investigate whether and how students reconciled conflicting ideas at the end of the relevant unit of the course. We examine these data sources to develop an initial model of interdisciplinary reconciliation in the context of ATP and chemical bond energy.

### *1. Multiple-choice quiz question*

Early in the second semester of the course, the students were given a quiz that included two multiple-choice questions taken directly from Galley [46] (given originally at the beginning of a physical chemistry course), for comparison. Here, we look at one of those questions:

> *An O-P bond in ATP is referred to as a "high-energy phosphate bond" because:*
>
> *A. The bond is a particularly stable bond.*
> *B. The bond is a relatively weak bond.*
> *C. Breaking the bond releases a significant quantity of energy.*
> *D. A relatively small quantity of energy is required to break the bond.*

Students were instructed to "put the letters corresponding to all the correct answers;" this is slightly different from Galley's students, who were given a limited set of choices ("A and C," "B and C," etc.). In both our class and Galley's class, choices B and D were considered the correct responses. The intent was that students would recognize that energy is released because a relatively strong bond was formed after a relatively weak bond was broken, and that no energy is released by the actual breaking of the bond.[3]

### *2. Interviews*

Over the course of the year, the research team conducted 22 semi-structured interviews with 11 students on various topics related to the course. Some of these interviews were designed as case studies to investigate how students were developing over time in this interdisciplinary course. In an effort to build in opportunities for triangulation with other data sources, these interviews often focused on specific course tasks. Semi-structured protocols were developed primarily to guide the interviewer in a set of research directions. A standard initial prompt was, "Have you encountered biology so far in this course? In what contexts?" These prompts were followed with probing questions to fully explore the contexts the students raised. At times this meant that a single prompt from the protocol guided the entire 45-minute interview.

Two of these interviews, with two pre-medical students, included explicit discussion of the ATP quiz question. The first interview, with Gregor[4], took place immediately after class on the day that the quizzes were handed back, and was the first interview completed with Gregor as a case study. Gregor brought up the quiz spontaneously in response to the prompt described above about the role of biology in the physics course. The second interview, with Wylie, was three weeks later and was the second interview completed with him as a case study. In the second interview, more time was spent on specific task prompts from the course. By this time, the research team had seen the Gregor interview data, so the interviewer prompted Wylie more directly about the ATP quiz question to explore how his reasoning compared with Gregor's.

[3] The question, as written, may have been misleading because it asks about the reason for using a term that is itself misleading. Because of this, we believe the question is more valuable as a formative task than as an assessment, and we focus on how the students subsequently thought through the question in interviews.

[4] All names are pseudonyms.

#### 3. In-class video

We collected video of the course for the entire year (embedding microphones with two student groups seated in different parts of the classroom). To investigate what contextual features of the pedagogy and curriculum may have supported reconciliation, we examine the directions that the instructor gave the students regarding the quiz, and a conversation between Gregor and the instructor immediately after the quizzes were handed back. These video data provide additional information on the larger classroom context, enabling us to understand the features of the course context that may have supported interdisciplinary reconciliation.

#### 4. Exam essay question

Using the data from the quiz, interviews, and in-class video we developed an essay question that would capture the ideas that students were grappling with in considering the energy in ATP. At the end of the thermodynamics unit, we administered this capstone essay question on a midterm exam, with the goal of observing and assessing interdisciplinary reconciliation for the whole class. All exams were scanned before returning them to the students, which allowed for further analysis after the exams had been handed back. A rubric for evaluating the ideas in the essay question was developed by a team of chemistry, biology, and physics faculty. The details of the question are discussed in section V.C.

### B. Data analysis

We use the four data sources in an interweaving way to address our research question. The fairly sparse multiple-choice data provided a baseline for how students were understanding ATP and chemical energy. Results from the multiple choice question and the in-class discussion inspired deeper probing of individual student reasoning through interviews. Finally, we spiraled back to understanding interdisciplinary reconciliation at the class level by developing an essay question reflecting the views from the individual student interviews. Similarly, understanding how we can characterize this interdisciplinary reconciliation as building upon existing theoretical frameworks leverages the details of reasoning in both the student interviews and the all-class essay question.

There are at least three methods that we could use to identify ideas articulated by students as belonging to particular disciplines: 1) the past experiences the students would have encountered (e.g. analysis of textbooks in the disciplines), 2) the ways in which disciplinary experts discuss the concepts, or 3) how the students themselves describe ideas as belonging to the disciplines. In this paper, we have chosen to examine interdisciplinary reasoning through the ways students label the disciplines, and as such we primarily focus on the ideas the students describe as belonging to physics and/or biology. We have shared these characterizations with our disciplinary colleagues to confirm that these descriptions are consonant with their disciplinary experiences, though we do not explicitly leverage these data in this work.

A significant component of our methodology in analyzing the data involves attending to the reasoning of individual students. In some sense, this limits the claims that we can make from the data, relative to a larger-N study. We do not claim that the results are directly generalizable to the entire student population. However, the case-study methodology allows us to examine the dynamics of individual student reasoning in ways that would be difficult to measure over a larger population [2]. In analyzing student reasoning, we focus our attention on the disciplinary context-dependence of students' modeling choices and on the process of reconciling apparently contradictory models associated with different disciplines.

## V. RESULTS

### A. Quiz and classroom data

On the quiz question, 79% of the class (N=19) selected choice C (breaking the bond releases energy) as a correct answer. Our sample size is too small to draw meaningful conclusions from a more detailed breakdown of the quantitative data. We bring up this result primarily to show that our class is broadly comparable to Galley's results, in which 87% of students chose C. Galley interprets this as a sign of a "persistent misconception." However, the qualitative data (which are the focus of the rest of this paper) illustrate that the picture is more complex than the multiple choice results suggest, and that our students are engaged in reconciling multiple disciplinary ideas, whether on their own or supported by the instructional context.

For initial insight into how disciplinary ideas are being reconciled, we look at interactions between the instructor and the students during lecture. This was the first quiz of the semester, and while the majority of the students in the class had been in the first semester of the sequence (with the same instructor), several students (including Gregor and Wylie) were new to the course. The instructor made a number of verbal moves, while administering the quiz and while returning it to the students, to attempt to reframe the meaning of the "quiz" activity in this course. While the quiz was being administered, the instructor emphasized the multiple-choice multiple-response format, in which students have to consider each answer option separately (rather than jumping to one "correct" choice as on a conventional multiple-choice question). He explicitly articulated his motivation for giving the students this format of questions: "Anything that I can do to undermine test-taking strategies and replace them with actual thinking, I will do." This may serve to communicate to students that they are intended to be thinking deeply about these questions, and not expected to quickly recognize or simply recall the correct answer.

On the next class day, when the quizzes were handed back, the instructor began the discussion by noting how little the quiz grades would contribute to the overall course grade. He explained that "the point of the quizzes is to get you thinking about stuff, and not so much as an evaluation of how well you are doing." Here, he attempted to renegotiate the purpose of "quizzes" with the students, explicitly describing them as formative opportunities to practice thinking, rather than as summative measures of students' success.

The instructor introduced the notion that students can argue for why a given quiz answer (which was marked wrong) should be considered correct. Discussing the ATP question, he said:

> *"The results were that 79% of you picked C. That's almost everybody. If you want to make an argument why you think that should be accepted as an answer, I will accept this as a regrade and consider, so write on the back, say 'I gave C, and I think you want to accept this as an answer because of the following.' If you have a good thoughtful answer … I don't know how they use the language in chemistry and biology, and if they talk about it as C, you might have a case. If you feel you can make it, do so. I'm perfectly willing to listen. That, by the way, is standard procedure here."*

Here we see the instructor describing how "correctness" on quizzes will be established. He shares with students that he will carefully listen to the arguments that they make and consider regrades based on the substance and sensibility of those arguments. Additionally, he acknowledges that different scientific disciplines may talk about these ideas in different ways. He also elaborates on these descriptions, and situates them as "normal" for the rest of the course.

There was some amount of discussion that followed the instructor's description of the quiz results. Gregor directly engaged with the instructor about how he (and, he surmises, other students in the class) were interpreting the question. This demonstrates that at least one student orients to the framing of the quizzes as an opportunity to engage about plausible interpretations of the quiz questions and answers:

> **Gregor:** *So I, and I guess probably a lot of other people, were assuming something that was not part of the question, that was that a more stable bond would be formed by breaking that bond, which would—*
> **Instructor:** *Write it up! Write it up. That's your justification.*
> **Gregor:** *But that's the assumption that we were making, is that—*
> **Instructor:** *Yes. And because you had some context that you were assuming that wasn't specified, or because this always happens in some context when you use it in biology that I wouldn't know about it, let me know. And if you convince me, I'll give you a point.*

In this interaction the instructor opens the door to other answers as being reasonable within a particular set of assumptions and says that, with those assumptions articulated and explained, he could be convinced that it is a reasonable answer. We also see here that the instructor points to the fact that these unarticulated assumptions may be associated with commonly used disciplinary contexts in biology. This conversation may have contributed to what Gregor then says in his interview.

## B. Interview data

Gregor had selected B, C, and D as correct answers on the quiz, and lost a point for choice C. In the interview immediately following the class when the quizzes were returned, Gregor responds to a prompt about the role of biology in the physics course, and explains why he chose C (though this retrospective explanation may or may not represent exactly what he was thinking while taking the quiz):

> *"I put that when the bond's broken that's energy releasing. Even though I know, if I really think about it, that obviously that's not an energy-releasing mechanism. Because like, you can't break a bond and release energy, like you always need to* *put energy in, even if it's like a really small amount of energy to break a bond. Yeah, but like. I guess that's the difference between like how a biologist is trained to think, in like a larger context and how physicists just focus on sort of one little thing. Whereas like, so I answered that it releases energy, but it releases energy because when an interaction with other molecules, like water, primarily, and then it creates like an inorganic phosphate molecule that has a lot of resonance. And is much more stable than the original ATP molecule. So like, in the end releases a lot of energy, but it does require like a really small input of energy to break that bond. So I was thinking that larger context of this reaction releases energy. Because I know what the reaction is, ya know? So, um, not, does the bond breaking release energy."*

Gregor demonstrates a sophisticated understanding of the ATP hydrolysis reaction, and makes clear that his justification for choosing C on the quiz does not correspond to the standard "misconception" that bond breaking releases energy. He displays understanding of the intended resolution of the apparent paradox: energy is released not by the breaking of a bond but by the formation of other more stable bonds. In thinking back over the question, Gregor stands by his answer, but also recognizes the correctness of the quiz answer key. He attributes the discrepancy to differing interpretations of what the question is asking (and even assigns this reasoning to the other students who answered the question the same way):

*"When I was taking the test, I guess I was thinking breaking this bond then leads to these other reactions inevitably. That result in an energy release … I don't [argue] that breaking a bond releases energy, but just like in a larger biological context, that reaction does release energy. So that's what me and apparently like 80% of the class was thinking."*

Gregor then, following up on a thread that begins above, ties these differences in perspective to differences between the disciplines:

*"Because I guess like in biology it's not as important to think about like breaking this bond doesn't release energy and then all these other things that happen do release a lot of energy. So, we're, I've just been taught like for a long time that like ATP going to ADP equals like a release of energy. … I guess that's just the difference between physics and chemistry and biology. … It's just your scale. Like, physic[ists] really love to think about things in vacuums, and like without context, in a lot of senses. So, you just think about like whatever small system you're—isolated system you're looking at, and I guess chemist or biologists thinking about more of like an overall context, that like wherever a reaction or process is happening, like that's important to what's going on."*

Gregor and his classmates have biology backgrounds, and their experience talking about ATP and bond breaking is in biology and chemistry courses; those experiences inform how he frames the context of the quiz question. Gregor now believes he is seeing a different perspective in a physics course, one in which the phenomenon of ATP hydrolysis is more narrowly conceived. He sees the boundary of the phenomenon under consideration as a salient difference between the disciplines. When Gregor says "scale," he is not talking about physical scale, but about whether we are looking at the breaking of a bond on its own (which requires an input of energy) or the ATP hydrolysis reaction as a whole (which releases energy).

Wylie also answered B, C, and D on the quiz. Like Gregor, both his multiple-choice responses on the quiz and his responses in the interview were consistent with holding two pictures simultaneously. However, Wylie apparently had not reconciled these two pictures prior to the interview to the same extent that Gregor had. The interviewer begins by handing Wylie a blank copy of the quiz on which the ATP question appeared. Wylie immediately affirms that he recalls the quiz from class and that he "picked something wrong," asserting that he answered option C, "for sure." As he is considering the other answers he had originally given on the quiz, the interviewer prompts him to say aloud what he is thinking. In the 10 minutes of discussion around the quiz question that follow, Wylie demonstrates awareness that he still has reconciliation to do. In thinking back over the question, he says "there's obviously a conflict" between breaking bonds (in ATP) releasing energy and forming bonds (in general) releasing energy. Wylie explains that he answered C because "the result of ATP hydrolysis is ADP, which is much more stable, because I know this from chemistry. … And we have energy released. So obviously you're going from an unstable state to a more stable state." He also justifies his choice of D ("a relatively small quantity of energy is required to break the bond"), "because if something is really unstable, if something is really highly charged, then all it needs is a little push, and that's it, it just goes downhill." Putting it together, Wylie says:

*"If I were to rationalize [the physics professor's] model, then I would have to say ATP breaks down into ADP plus something. There's a bond formed between the phosphate and something that makes it more stable. And this part is the part that releases the energy. … It's not the breaking of*

*the bond that's releasing the energy. Because when, in breaking of the bond, you actually require energy, but the result of the breaking of the bond is that you get energy."*

Even though Wylie does this reconciliation, possibly in real time during the interview, to explain why C was deemed incorrect, he remains unsatisfied with this conclusion. Like Gregor, Wylie ultimately connects the discrepancy to disciplinary differences:

> **Wylie:** *If … that same question was in a biology course, and I picked C, I would get points.*
> **Interviewer:** *Why do you think that is?*
> **Wylie:** *Because I think in the biology course, the focus of the question would be on the significant quantity of energy, not necessarily breaking the bond. … Breaking the bond in ATP gives you energy. That's what a biologist might think. … But this is more specific. This is going into, you know, exact details.*

Wylie, too, distinguishes between a "biology" approach, in which the focus is on the entire reaction that releases energy, and the "more specific" approach that he associates with the physics class, which focuses on the "exact details."

At the end of the day, Wylie has not gone all the way in building a coherent model and knows that he has further to go:

> *"But … I keep thinking that there have to be things that, you know, just like ATP, you know they are macromolecules. They're not as stable together, but when they break down they're more stable separately...what do you do with that? How would you release energy in that sense? I don't know. I'm really just kind of unclear on that."*

## C. Exam essay question

In reviewing the interview data from Gregor and Wylie we see the two sets of ideas around ATP being clearly articulated, but with distinctions in disciplinary ideas being drawn between them. We used the clear articulation of these ideas from the interviews to form a capstone assessment question. This question drew students' attention to the difference between focusing exclusively on the breaking of the O-P bond in the ATP molecule and allowing the system under consideration to include the forming of bonds with the H2O molecules surrounding the ATP, just as Gregor and Wylie have articulated. The capstone assessment was given as an essay question on the first exam in the second semester of the course. The essay question (shown in Figure 1) presents two contrasting arguments, allowing for other students in the class to demonstrate their understanding of the reconciliation of these ideas. Based on the ways that students associate the disciplines with these arguments in interviews, we see the focus exclusively on breaking the O-P bonds as associated in the students' views with "physics" and the focus on the entire ATP hydrolysis process as associated with "biology."

**4. (10 points)** Two students discussing the process of ATP hydrolysis (ATP + $H_2O$ → ADP + $P_i$) make the following comments: **Justin**: "The O-P bond in ATP is called a 'high-energy bond' because the energy *released* when ATP is hydrolyzed is large. That released energy can be used to do useful things in the body that require energy, like making a muscle contract." **Kim**: "I thought chemical bonds like the O-P bond in ATP could be modeled by a potential energy curve like this (she draws the picture at the right), where *r* is the distance between the O and the P. If that's the case, then breaking the O-P bond in ATP would require me to *input* energy. I might not have to input *much* energy to break it, if that O-P happens to be a weak bond, but shouldn't I have to input at least *some* energy?" How did Kim infer from the PE graph that breaking the O-P bond requires an input of energy? Who's right? Or can you reconcile their statements? (The chemical structures of this process are given if you find that useful.) *Note: This is an essay question. Your answer will be judged not solely on its correctness, but for its depth, coherence, and clarity.*

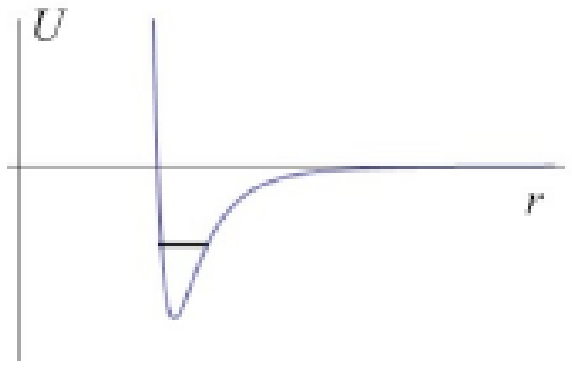

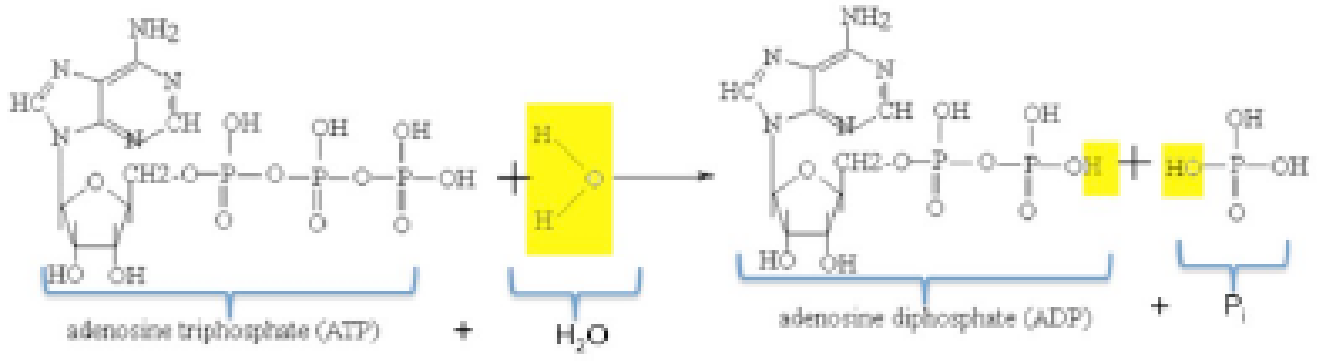

**Figure 1.** The reconciliation essay question given to students on the midterm exam.

In the essay question Justin articulates that hydrolyzing ATP releases a large quantity of energy, which is why the O-P bond in ATP is typically called a “high-energy bond.” In contrast, Kim describes the O-P bond with a potential-energy curve that shows a negative energy. Kim reasons that getting out of this negative energy well (which represents the O and P moving far apart from each other) would require an input of energy. The essay question explicitly asks students to make sense of the potential-energy curve and use it to explain why Kim’s response makes sense. Then the essay prompts students to take a position as to whether Justin or Kim is correct, or to reconcile the two statements.

Kim’s and Justin’s statements were juxtaposed in this essay question to explicitly draw attention to the differences in reasoning. Students in the class were accustomed to this type of essay question, in which two ideas are presented and the expectation is that the students take up one side or the other of the argument. However, in the case of this prompt, both Kim and Justin are explaining correct ideas. The intention with this capstone assessment was to examine the extent to which the students in the class attempted to reconcile these perspectives and to examine the form of that interdisciplinary reconciliation.

Student responses to the essay question varied widely, from some who said plainly that Justin’s idea about energy being released when ATP is hydrolyzed is wrong, to others explaining quite articulately why the reasoning from both Justin and Kim are correct. Essays from Sameer and Sebastian represent what we consider to be exemplary responses demonstrating interdisciplinary reconciliation.

> **Sameer:** *In general, Kim is correct. Justin is correct in that ATP is a vital source of energy for biological reactions, but he’s confusing the energy of ATP hydrolysis with the energy of the bond. By the principles of thermodynamics (illustrated by the chart above), atoms do want to be bound together, as it represents the lowest potential energy level. Due to their attractive forces, it requires energy to break the bond – the stronger the bond, the lower the energy state, and the greater the amount of energy required to break the bond.*
>
> *With ATP, the O-P bond has a higher energy state, and does not require as much energy to leave the bound state. When new bonds form, though, they are at a much lower energy state than the O-P bond (deeper well), so large amounts of energy are released to the environment. It is not the bond breakage that releases energy, but rather the formation of new bonds, and this is where the confusion lies.*

> **Sebastian:** *They are both right. On the graph, at z* [z is indicated on the PE curve at the location of the horizontal line showing total energy] *the atoms are bonded and they are at a lower potential energy which is where they want to be. In order to increase r (the distance b/w the atoms) you would have to put in energy to increase the energy. Makes sense. So Kim is right that even though it is a weak bond you still have to put in some energy. Justin is also right because the formation of ADP releases more energy than required to break the initial O-P bond. Therefore there is a net release of energy. Just like in the graph you lose PE when the molecules come together rather, it is released. Justin is right too.*

In Sameer’s and Sebastian’s responses we see evidence that the students see the correct ideas within the two apparently contradictory statements that breaking the O-P bond requires an input of energy and that ATP hydrolysis results in a net output of energy. Both Sameer and Sebastian emphasize the formation of new bonds in describing the hydrolysis process and Sameer even goes so far as to explain that “this is where the confusion lies.”

This is not to say that all the students in the class demonstrated this understanding of the reconciliation this task afforded for students. Ava and Zeke demonstrate an alternative but relatively common response to the question, which favors the reasoning from Kim:

> **Ava:** *Kim says that breaking the O-P-bond requires an input of energy because as shown at the black line, that is the total energy that the O-P bond has, since they are bound together. Being in a bound state means that you are at a lower potential energy, such as the dip in the graph. Therefore, to break the bond, or to get out of that dip, you would need to input the amount of energy that is ≥ than the negative total energy that the molecule has. Therefore, Kim is right. Breaking a bond needs at least some energy because the O-P bond* <u>*is*</u> *a weak bond. Justin is not right because energy is not released when ATP is hydrolyzed. He is mistaking the phrase "high energy" in that there is no high energy in the bonds.*

> **Zeke:** *I think Kim is right. The energy released when the O-P bond broken is NOT large like Justin says (when ATP is hydrolyzed). It’s called a high energy bond because it’s a weak bond and relatively easy to break. Kim inferred that energy is required to break the bond since the potential energy is negative when the O-P are in a bound state. So she feels that she must input that amount of energy to make them unbound. I think Kim is* *right and her graph makes sense. You would need to put in the amount of energy that is shown as –U on the graph to break the bond. Justin is right in saying that the energy that is then released can be used to do useful work. So you have to input energy to break it, but more useful energy is released by breaking it.*

In both Ava's and Zeke's responses we see an emphasis on describing how Kim is right because breaking the O-P bond requires an input of energy in order to get out of the potential energy well. However, in Ava's response we see an explicit reference to Justin's ideas being incorrect because she concludes (incorrectly) that energy is not released when ATP is hydrolyzed. In Zeke's response there is evidence of the reconciliation not quite being worked out. He clearly describes the O-P bond requiring energy to break, though it is a relatively weak bond. However, when he describes Justin being right, it is unclear whether he thinks both that it requires energy to break the bond and that breaking the bond releases energy.

We note a difference between the multiple-choice quiz question discussed in the interviews and this exam essay question: Both Justin and Kim in the essay question present scientifically correct ideas. It does require energy to break the O-P bond and to move out of the negative potential-energy well. At the same time, ATP hydrolysis, which involves the breaking of the O-P bond as well as the forming of new bonds between the phosphate molecule and the water surrounding the molecule, does result in a net energy release. This is different from the more common type of reconciliation task, asking students to abandon an "incorrect" idea in favor of the more scientifically "correct" one, which is what is required by the quiz question presented earlier.

We believe that the essay question provided students with the opportunity to reconcile these ideas, as nearly half of our 19 students in response to this prompt discussed both the forming and the breaking of bonds, which have different implications for the net energy effects. We claim that these students demonstrate recognition that both of these ideas hold some value to making sense of the language of "high-energy bond" surrounding ATP hydrolysis, and that by examining the essay responses in detail we have gained insight into how students reconcile these ideas. When students seek consistency between Kim's and Justin's ideas, we see them as grappling with the relationship between disciplinary perspectives.

# VI. DISCUSSION

## A. Interdisciplinary reconciliation

In this paper we have developed a model for the process of interdisciplinary reconciliation (IDR) that draws from classroom supports, fine-grained student reasoning in interview contexts, and broad-stroke data from student essay responses. We draw attention in this initial model to both the endpoint that Gregor exemplifies in the interview setting as well as the groundwork for this target in the instructional supports.

### *1. The target endpoint for IDR*

Gregor's interview response represents what we see as an exemplary aspect of interdisciplinary reconciliation:

*Physic[ists] really love to think about things in vacuums, and like without context, in a lot of senses. So, you just think about like whatever small system you're—isolated system you're looking at, and I guess chemist or biologists thinking about more of like an overall context, that like wherever a reaction or process is happening, like that's important to what's going on.*

In Gregor's words we see a sophisticated understanding of the modeling choices he has encountered in physics and biology. In the interview, Gregor ties these choices to the reasoning behind the different answers about the O-P bond in ATP. He connects his answers about the energy released in ATP hydrolysis to the kinds of questions one might ask in the different disciplines and the different ways the disciplines define the boundaries of systems. We see in Gregor's response a compelling example of the place where students get to in the process of interdisciplinary reconciliation.

In interdisciplinary reconciliation (IDR), both disciplines represent locally coherent sets of canonically correct scientific ideas that can be activated for different purposes. While some biologists would take issue with Gregor and Wylie's "biology" statements, our conversations with biologists have corroborated Wylie's claim that his answer would be considered correct in many biology instructional contexts. More importantly, the idea that ATP hydrolysis results in a net release of energy is particularly useful for reasoning in biological contexts. The goal, then, is for students to understand the connections between disciplinary ideas while maintaining each one in its appropriate context.

In addition to achieving conceptual reconciliation, an important outcome of interdisciplinary reconciliation is that students will be able to activate the appropriate disciplinary idea(s) depending on the context. Therefore, part of the reconciliation is developing the resources to distinguish contexts, to establish the appropriate framing in each context, and to understand why an approach is most productive in a given context.

### 2. Instructional components of IDR

In addition to the endpoints displayed in the interview and essay data, we also consider the elements of the instructional environment that contributed to the process of IDR. In the classroom video data, the instructor continually attempted to engage students in rethinking the purpose of activities, encouraging them to think carefully and consider all the options available to them, which may have included disciplinary ideas. The instructor set up quiz questions as multiple-choice multiple-response, which is in direct contrast with common multiple-choice questions where there is only a single correct answer. Multiple-choice, multiple-response questions allow students to consider the possibility that there can be multiple ideas that are correct. This style of quiz lends itself to exploring disciplinary ideas, which may have different languages and assumptions associated with similar (correct) ideas.

The instructor in discussions and other elements of the course also tried to communicate that disciplinary language and starting assumptions might not be transparent, and that students could help in clarifying these disciplinary ideas. As mentioned in section III.B, the course was structured to encourage students to bring in ideas from the various scientific disciplines. One example can be seen in the prompts students encountered in the class. Both the quiz prompt and the essay question described above explicitly bring up ATP and ideas of stability and weakness of the O-P bond. These ideas are more commonly encountered in biology or biochemistry classes, and by situating them within a physics classroom we encourage students to examine the differences in the disciplinary descriptions of these ideas.

Similarly, when the instructor discussed the quiz results with the students, he directly referenced the potentially different uses of language in chemistry and biology, and encouraged students to consider that language when making arguments about points deserved on the quiz. Further, the instructor acknowledged that particular assumptions and ways of bounding the system that the students might be using are more common in biology contexts. Through the explicit references to scientific disciplines and traditional disciplinary ideas, the course and the instructor communicated to students that some ways of reasoning through the phenomenon might be discipline-specific and not unique.

We conjecture that the reasoning displayed by the students in the interviews and in their responses to the essay question was encouraged by these supports from the course and instructor. In identifying a model for IDR we draw attention to these instructional moves as important aspects of the process of reconciling disciplinary ideas.

## B. IDR and other frameworks for reconciliation of ideas

Our primary research question asks how existing theoretical frameworks can help characterize interdisciplinary reconciliation. In this section and the next, we compare and contrast IDR to existing frameworks.

An essential part of understanding student learning with a focus on conceptual ideas is having a model of what happens when students change their ideas. In particular, research on learning and conceptual change has focused on what students do with two sets of ideas that on the outset appear to be in direct conflict with one another. We focus on two ways of creating instructional activities to deal with this situation that have received attention in the physics education research literature: *elicit-confront-resolve* (*e-c-r*) and what is commonly known as *Elby pairs*.

Both of these processes for reconciliation begin with the same basic step: elicit commonly held student beliefs that are typically in conflict with widely held scientific concepts. The second step centers around confronting those beliefs with experimental data that is in conflict with the originally elicited belief. The difference in the two methods lies in the process of reconciliation. We describe each of these methods in more detail below.

### 1. Elicit-confront-resolve

The *e-c-r* method focuses on replacing an old unscientific idea with a new one that is more aligned with those that are held to be scientifically correct. The University of Washington Physics Education Group has spent a large part of their efforts identifying and documenting "student difficulties" – sets of conceptual ideas that students bring to an introductory physics environment that can be reliably activated with a particular set of contexts and questions [1,55]. In developing highly successful curricular reforms, they have identified a process they call *elicit-confront-resolve*. The strategy begins with a well-structured question that elicits students' commonly held beliefs. Next, that belief is confronted with contradictory experimental evidence that the student observes individually. Finally, in the *resolve* step of the process, students practice applying physics rules in carefully structured examples that help them to overcome their tendency to apply commonly held beliefs, and replace them with the scientifically correct conceptions.

While the elicit-confront-resolve framework has been highly successful in achieving student understanding of complex ideas within physics, one of the fundamental requirements of this instructional strategy is the idea of replacing the student's naïve conception with a new scientifically correct conception. This replacement is problematic when the conflicting ideas to be resolved are those from different disciplines, and neither is more "correct" than the other. In the elicit-confront-resolve framework, we can easily imagine many students giving similar reasoning to Ava's exam response where she says clearly, "Justin is not right because energy is not released when ATP is hydrolyzed." From all of her biology experience, Ava must know that the hydrolysis of ATP does release energy, but here we see her going past the point of reconciliation. Indeed she appears to have resolved the conflict, but in doing so has abandoned her (correct) biology knowledge in favor of using correct physics reasoning about breaking bonds. This kind of reconciliation is not consistent with our goals for an interdisciplinary environment.

### 2. Elby pairs

An alternative pathway has been proposed by Elby [5], reasoning from the resources framework. The framework, as we described in section II, allows for students holding multiple ideas at the same time, and for the activation of particular ideas to be related to the context at hand. Elby pairs (a term coined by Redish [56] to describe a reconciliation process outlined by Elby [5]), are a set of two questions that ask the same question in two different ways. The first way is designed to elicit the common idea in a similar way as the *elicit* component of the *e-c-r* process. However, the second question is designed to match the students' correct intuition about the situation. The classic example given in both Refs. 5 and 56 centers around Newton's third law. One question in the Elby pair elicits students' intuition that when a truck collides with a car, the truck exerts a larger force on the car than the car does on the truck. The second question activates students' correct sense that the momentum changes of the interacting objects are balanced. Reconciling their answers to the pair of questions leads students to conclude that the car **accelerates** more than the truck, but also has a smaller mass, so their intuition that the car is affected more than the truck is in fact consistent with Newton's third law.

The key difference between the Elby pairs version of reconciliation and the *e-c-r* version is that the Elby pairs are designed to invite students to see the pair of questions as refinements of the same raw intuition. Thus the Elby pairs version of reconciliation does not ask students to replace one idea with the other, but rather to recognize how their initial ideas need to be refined in order to be aligned with the scientifically correct assumptions. In so doing, students are not asked to abandon their initial set of ideas, but rather to modify them in order to resolve the conflict.

Seeking to support students in refining their raw intuition to develop an understanding of how their intuition can be productive in a physics class is a positive and encouraging goal. However, in an interdisciplinary environment, we are not dealing with students who need to refine a raw intuition into a more scientifically correct conception. Instead, these students may already have productive ideas that have come from scientifically correct ideas in biology and chemistry contexts. Our goals for reconciliation should then be to encourage maintaining both sets of conceptions with a deep understanding of the assumptions that underlie them and the context of their utility: a vision that was reached through careful examination of exemplary student reasoning.

### 3. Interdisciplinary reconciliation

We argue that the model of interdisciplinary reconciliation that we present in this paper is different from both the *e-c-r* and Elby pairs types of reconciliation. As discussed above, both disciplines involved in IDR represent locally coherent sets of canonically correct scientific ideas. Therefore, the goal of IDR is not to arrive at one refined idea (a common goal of both *e-c-r* and Elby pairs), but to refine both ideas to the point where students can understand how the two disciplinary ideas fit together. In Elby pairs, both explanations are shown to be based in the same raw intuition; in IDR, the two disciplinary ideas may come from different places, and the role of reconciliation is to bring them into coherence.

While much of the previous work on context-dependent activation of resources can be interpreted as operating under the assumption that one pattern of resource activation is correct in answering a given question, we take the position for interdisciplinary questions that disciplinary context-dependence is productive and is one of the goals of IDR.

Of course, it is not always the case that when students are reconciling multiple ideas, both ideas are correct. Therefore, the other approaches to reconciliation are still appropriate in many circumstances. We describe IDR as appropriate in a limited set of cases, when the ideas to be reconciled are correct ideas from different disciplines, but this set of cases is increasingly significant with the development of more interdisciplinary curricula.

## C. Misconceptions and resources, revisited

We introduced the misconceptions and resources frameworks in section II, and we now apply them to our data to compare our approach with previous work on student learning of the same scientific phenomena. In analyzing student difficulties with ATP, both Novick and Galley [42,46] invoke the misconceptions model. Novick writes, "many students conceive chemical energy as stored in something called a chemical bond … Now this is obviously a serious scientific misconception." To remedy this, he suggests eliminating the energy storage language from textbooks. Galley writes that students "adhere to the belief that energy is obtained when chemical bonds are broken," and attributes this primarily to "misinformation" in biology courses. He concludes, "[i]f students are alerted to the confusion and misinformation about bond making and bond breaking that they were likely exposed to, coupled with a review of the correct picture of bond rupture and formation, the problem is largely resolved. Students then recognize the misconceptions that they encounter." In both cases, the model of students is that they hold only one view about chemical bonds, and if the source of this incorrect view is eliminated or confronted, then students will replace it with the correct picture.

We can make some predictions from a misconceptions model. If students have a unitary belief that energy is stored in bonds, we would then expect them to say that a stronger bond is a bond that stores more energy (so that more energy is released when such a bond is broken), and to reject the idea that an input of energy is required to break bonds. After they are convinced of the correct view of bond breaking and bond formation, we would expect them to abandon their previously held incorrect view. The misconceptions perspective would suggest a vision for interdisciplinary education in which one view is eliminated and one specific view is maintained.

Instead, our data are more consistent with the resources framework. On the ATP quiz question, several students answered both C and D: that breaking the O-P bond releases a significant quantity of energy and that breaking the bond requires a small amount of energy. Other students answered both B and C: that the bond is relatively weak and that it releases a large quantity of energy. Many instructors would see these combined responses as mutually inconsistent, suggesting that breaking the O-P bond both requires an input of energy and releases energy or that the bond is weak and releases a large amount of energy. The fact that many students did supply these two apparently inconsistent ideas is difficult to explain using a unitary "misconceptions" view. However, from a resources perspective we can view these apparently contradictory ideas as being sets of activated resources.

Our interview data with Gregor and Wylie suggest that the disciplinary context has a key role in determining which resources are activated. When Gregor explains his reasoning,

> *I guess that's the difference between like how a biologist is trained to think, in like a larger context and how physicists just focus on sort of one little thing. Whereas like, so I answered that it releases energy, but it releases energy because when an interaction with other molecules, like water, primarily, and then it creates like an inorganic phosphate molecule that has a lot of resonance…. So like, in the end releases a lot of energy, but it does require like a really small input of energy to break that bond.*

he explains how his biology training encouraged him to consider the ATP molecule's interaction with other molecules around it, such as water. This explanation suggests that Gregor has access to both the resource that says energy is required to break the O-P bond and the resource that explains the large amount of energy released from ATP hydrolysis, and the disciplinary context created the framing that activated the set of resources he used to answer the question. Furthermore, Gregor demonstrates an awareness of his access to these different resources, suggesting that one of them is not more likely to be accessed than the other unless prompted by a disciplinary context. Wylie makes an explicit reference to the disciplinary context deciding which resource is the most appropriate to bring to bear when he says,

> *Because I think in the biology course, the focus of the question would be on the significant quantity of energy, not necessarily breaking the bond. … Breaking the bond in ATP gives you energy. That's what a biologist might think.*

We do not mean to suggest that all students may be as self-aware as Gregor and Wylie in articulating the connection between the sets of ideas that should be brought to bear and the disciplines. However, we do believe that the responses from Gregor and Wylie highlight the advantages of a resources framework over a misconceptions view in explaining these data. The dynamic view of the resources framework allows us to make sense of how students could be articulating apparently contradictory ideas within the same set of responses, as well as how the disciplinary context may change which ideas students bring to bear.

# VII. CONCLUSIONS

In this paper, we demonstrate that when biology students encounter ATP in a physics course, reasoning about chemical bond energy in an interdisciplinary context is a complex process requiring students to manage ideas that may seem contradictory on the surface. We provide examples of what interdisciplinary reconciliation looks like in the context of ATP hydrolysis and highlight the seemingly contradictory scientifically correct ideas that students must learn to navigate. We explain this reasoning process within a resources view of student cognition, in which students' ideas are not unitary and coherent, but can be fragmented and dynamic. This model of student learning helps us build an understanding of how students can display coherent reasoning with two seemingly contradictory ideas, and how those ideas may be encouraged through disciplinary contexts. We present an alternative view of reconciling student ideas that embraces disciplinary differences, and encourages students to make explicit the assumptions that may be behind particular disciplinary reasoning. Finally, we point to particular instructor and course supports that may have encouraged the interdisciplinary reconciliation process.

The *Vision and Change* report calls for future biologists to develop expertise in another scientific discipline and to "develop the vocabulary of both disciplines and an ability to think independently and creatively in each as well."[29] We share this vision of interdisciplinary education, which does not suggest eradicating disciplinary differences. Instead, this vision emphasizes being able to reason **within** each discipline, using its own native tools, in ways that are informed by and coherent with the other disciplines.

We want our students to be able to make choices about how to model a system or phenomenon based on the questions that they are trying to answer. In some circumstances, it is appropriate to consider the individual steps of the ATP hydrolysis reaction mechanism and keep track of which bonds are broken and which bonds are formed, or to track the energy transformations and transfers that take place within this reaction. In other circumstances, the aqueous environment is backgrounded. The relevant features of the reaction are that ATP is broken into ADP and phosphate and that energy is released, and this relatively black-boxed picture is a useful way to think about the reaction in its larger biological context. Interdisciplinary competency in physics, biology, and chemistry incorporates both of these models, as well as the flexibility to move coherently among the models. In this disciplinary context-dependence we see the roots of productive interdisciplinary reasoning.

Future directions for research include applying the interdisciplinary reconciliation framework outlined in this paper to other content areas. This work has begun with an analysis of students' interdisciplinary reasoning about entropy, free energy, and spontaneity.[57] We have also identified cases in which "interdisciplinary" reconciliation is appropriate even within a single discipline, when different conceptual and epistemological resources are called for in different subfields. For example, at the level of professional physics, the modeling choices made in condensed matter physics and in particle physics are very different. In the standard introductory physics course, the simplifying assumptions made about energy are very different when energy is encountered in the contexts of mechanics and thermodynamics. A future analysis can explore the similarities and differences between this sort of intradisciplinary reconciliation and the interdisciplinary reconciliation discussed in this paper.

# ACKNOWLEDGMENTS

The authors thank Ben Geller and Arnaldo Vaz for valuable contributions to the thinking represented in this article. We are particularly grateful to Chris Bauer, Melanie Cooper, Catherine Crouch, and Mike Klymkowsky for their aid in developing the rubric for evaluating the essay question, and thinking critically about the energy in ATP hydrolysis, as well as providing insight into disciplinary perspectives from biology, chemistry, and physics. We would also like to thank the University of Maryland Physics Education Research Group and Biology Education Research Group for their valuable feedback on the arguments presented here. This work is supported by the NSF Graduate Research Fellowship (DGE 0750616), NSF-TUES DUE 11-22818, and the Howard Hughes Medical Institute NEXUS grant. Any opinions, findings, and conclusions or recommendations expressed in this publication are those of the authors and do not necessarily reflect the views of the National Science Foundation or the Howard Hughes Medical Institute.